\documentclass[12pt]{article}
\usepackage{epsfig}
\usepackage{float}
\usepackage{subcaption}
\usepackage{amsmath}

\def\beq{\begin{equation}}
\def\eeq#1{\label{#1}\end{equation}}
\def\eeqn{\end{equation}}

\def\beqa{\begin{eqnarray}}
\def\eeqa#1{\label{#1}\end{eqnarray}}
\def\eeqan{\end{eqnarray}}

\let\bar=\overbar

\def\Dslash{\not{\hbox{\kern-4pt $D$}}}
\def\dslash{\not{\hbox{\kern-2pt $\del$}}}

\def\msb{{\bar{\ssstyle M \kern -1pt S}}}

\def\Title#1{\begin{center} {\Large {\bf #1} } \end{center}}

\begin{document}

\Title{Kaon-Scatter Introduced Backgrounds in the KOTO Experiment}

\bigskip\bigskip


\begin{raggedright}  

{\it Stephanie~Su\index{S. Su.}\\
Department of Physics\\
University of Michigan\\
Ann Arbor, MI, USA}
\bigskip\bigskip
\end{raggedright}

\section{Introduction}
\begin{figure}[htb]
\begin{center}
\epsfig{file=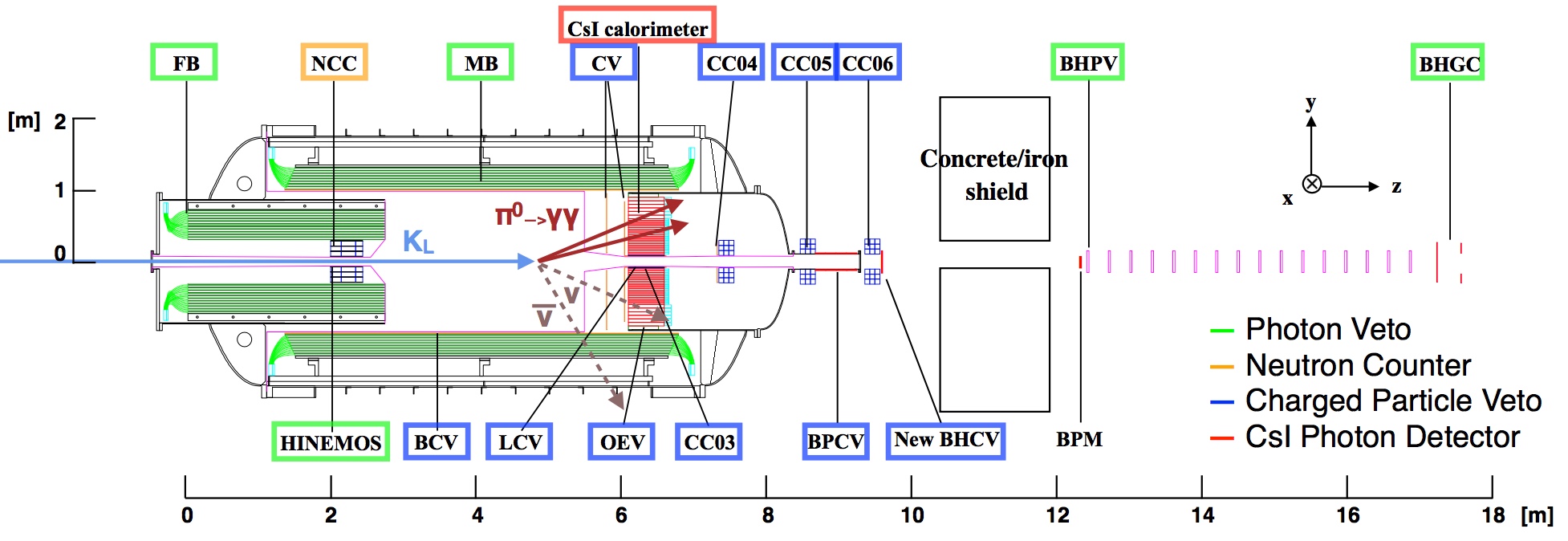,width=\textwidth}
\caption{Cross-view of the KOTO detector. The Cesium Iodide (CsI) crystal calorimeter is surrounded by many veto detectors, which includes scintillation detectors, cherenkov detectors, wire chambers, and crystal detectors. The green lines suggest that the detector is a scintillation detector.}\label{fig:detector}
\end{center}
\end{figure}

The KOTO experiment is a particle physics experiment located at the J-PARC facility in T$\overline{\mbox{o}}$kai-mura, Ibaraki, Japan. The goal is to measure the branching ratio of the direct CP-violating decay $K_L\rightarrow\pi^0\nu\bar{\nu}$, whose branching ratio is calculated to be $(3.00 \pm 0.30)\times 10^{-11}$\cite{BR} in the Standard Model (SM). the current experimental upper limit established by KEK E391a is $2.6\times10^{-8}$\cite{E391a}. This high precision theoretical calculation for the branching ratio enhances the sensitivity towards probing for new physics beyond the SM. The KOTO experiment receives a secondary $K_L$ neutral beam generated by a 30~GeV primary proton beam hitting on the gold target. The proton beam is slowly extracted in 2~seconds, with 6~seconds repetition. The decay of $K_L\rightarrow\pi^0\nu\bar{\nu}$ is measured by observing the two photons generated by a single $\pi^0$ in the calorimeter and nothing in the other detectors. The KOTO detectors consist of a Cesium Iodide (CsI) calorimeter surrounded by veto detectors sensitive to photons, neutrons, and charged particles, as shown in Fig. \ref{fig:detector}.

\section{$K_L\rightarrow\pi^0\nu\bar{\nu}$ Background}

\begin{figure}[h]
    \centering
    \begin{subfigure}[t]{0.5\textwidth}
        \centering
        \epsfig{file=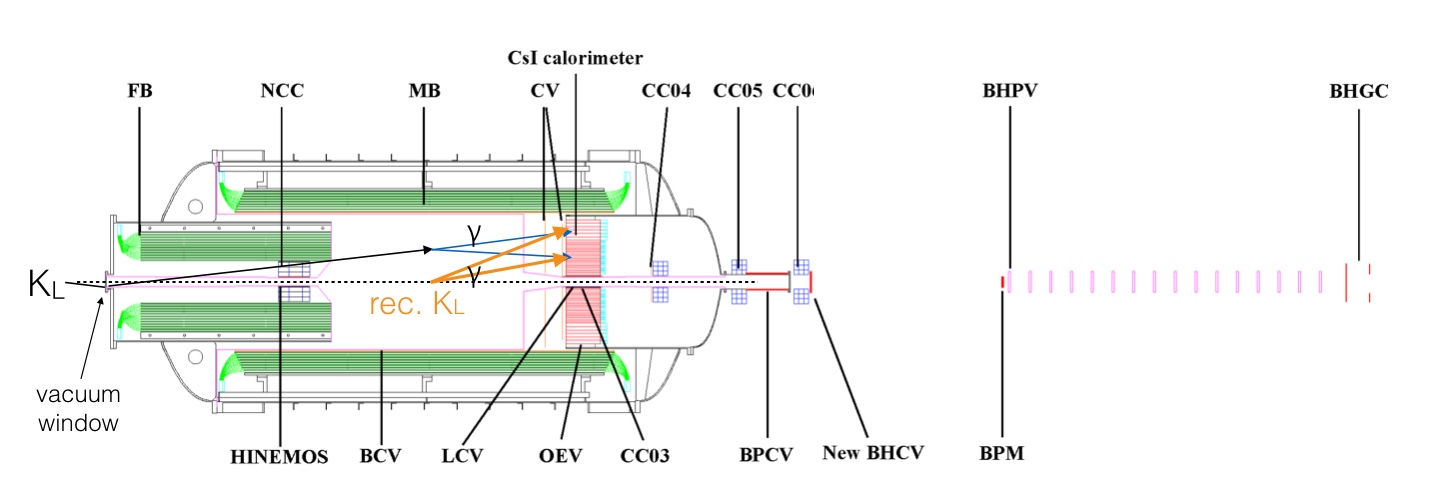,width=\textwidth}\label{fig:KL2g}
    \end{subfigure}%
    ~ 
    \begin{subfigure}[t]{0.5\textwidth}
        \centering
         \epsfig{file=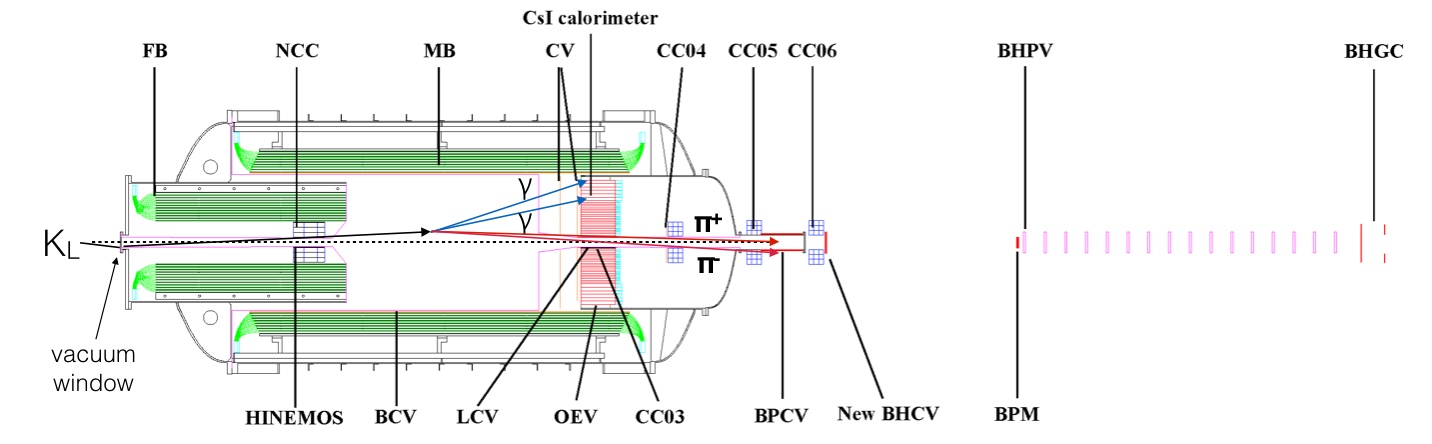,width=\textwidth}\label{fig:KLpipipi0}
    \end{subfigure}
    \caption{$K_L$ scattered at the vacuum window and are reconstructed back to the beam axis, thus to have appearent large transverse momemtum.}\label{fig:KLscatter}
\end{figure}

The signal of $K_L\rightarrow\pi^{0}\nu\bar{\nu}$ decay has the signature of final product with two photons on the CsI calorimeter with no signal on the veto detectors. It also possesses a large transverse momentum due to the missing neutrinos. Having the veto detectors wrapped around the calorimeter allow us to eliminate most of the background decays from our signal. However, kaons that decay outside the beam line with final product of two photons, such as $K_L\rightarrow\gamma\gamma$ and $K_L\rightarrow\pi^{+}\pi^{-}\pi^{0}$, can appear to have large transverse momentum due to kaon scattering at the vacuum window. The current reconstruction method uses assume all the $K_L$ come from the beam axis. Hence, when the off-axis $K_L$ decays, it will be reconstructed back to the beam axis and appear to have a large transverse momentum.  These off-axis kaon decay events can impact the measurement of the upper limit of $K_L\rightarrow\pi^{0}\nu\bar{\nu}$ branching ratio. Figure \ref{fig:KLscatter} illustrates different situations of the decays from the scattered $K_L$ at the vacuum window.

\section{Scattered $K_L$ Study}
\begin{figure}[htb]
\begin{center}
\epsfig{file=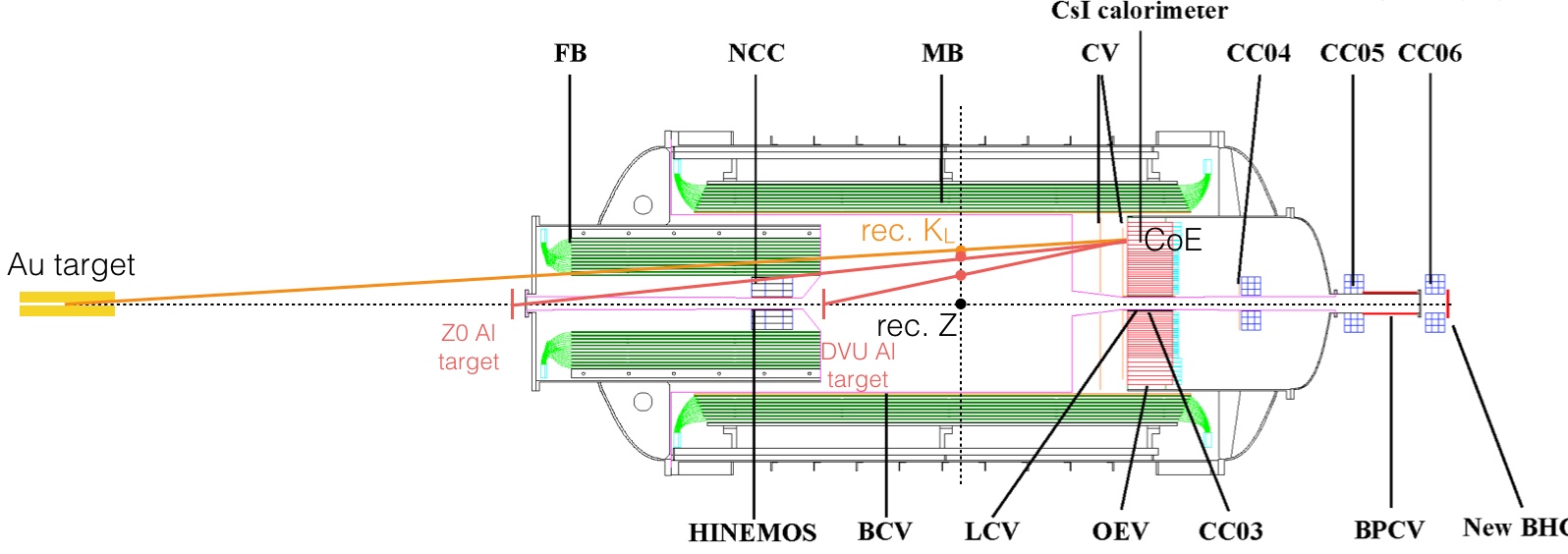,width=\textwidth}
\caption{$K_L$ vertex reconstruction method. The $K_L$ decay position is determined by the projection of the CoE to the gold target, which leads to larger reconstucted $K_L$ postion (yellow dot) compared to the acutal $K_L$ decay position (red dots).}\label{fig:scatter}
\end{center}
\end{figure}

The current reconstruction method first uses the $\pi^0$ mass as constraint and use the photon energy and position to determine the decay vertex, then release the $\pi^0$ mass constraint and use the found decay vertex as a new constraint to determine the $\pi^0$ mass. Finally, assuming all $K_L$ came from the gold target, we project the Center-of-Energy (CoE), which is another good indicator of the magnitude of the transverse momentum, back to the gold target to obtain the $K_L$ decay x and y positions. An illustration is shown in Fig. $\ref{fig:scatter}$. The CoE radius is defined to be $\frac{\sqrt{(\sum_i E_ix_i)^2+(\sum_i E_iy_i)^2}}{\sum_i E_i}$ and detailed information can be found in the reference\cite{SuIEEE}. For a scattered $K_L$ decay, the reconstructed $K_L$ vertex will have a more wild-spread beam profile distribution, as well as apparent larger transverse momentum.\\

\noindent In the KOTO experiment, there are two removable aluminum (Al) targets (Z0 Al target and DVU Al target) located upstream of the KOTO detector and inside the decay-volume, which is important towards the study of scattered $K_L$ decays. With the Al target inserted, we can operate the same reconstruction method to obtain the accurate decay vertex since the $K_L$ generation point is determined by the location of the Al target and there is no scattering or other interactions of $K_L$ happened inside the decay volume. A known decay vertex can be determined by studying the $K_L\rightarrow\pi^0\pi^0\pi^0$ decay with the Al target inserted, given the decay has enough constraints. The study is proceeded with the Monte Carlo (MC) study and with the data collected in the 2017 June run. By studying the scattered $K_L$ from the Al targets, we can scale the effect by the thickness and the density of the Al target to the Kapton-made vacuum window. The background estimation of the scattered $K_L$ at the vacuum window can be made by scaling down from the properties of the Al target to the material property of the vacuum window, as shown in Eq. \ref{eq:scale}, where the thickness of the Al target and the vacuum window is 5~mm and 0.125~mm respectively, and the density is 2.699~g/cm$^3$ and 1.42~g/cm$^3$ respectively.

\begin{align} \label{eq:scale}
\mbox{scatter K}_{\mbox{L}\, _{\mbox{vac}}} &= \mbox{scatter K}_{\mbox{L}}\, _{\mbox{Al}} \times \frac{\mbox{thickness}\,_{\mbox{vac}}}{\mbox{thickness}\,_{\mbox{Al}}} 
\end{align}

\subsection{Monte Carlo Study}
\begin{figure}[htb]
\begin{center}
\epsfig{file=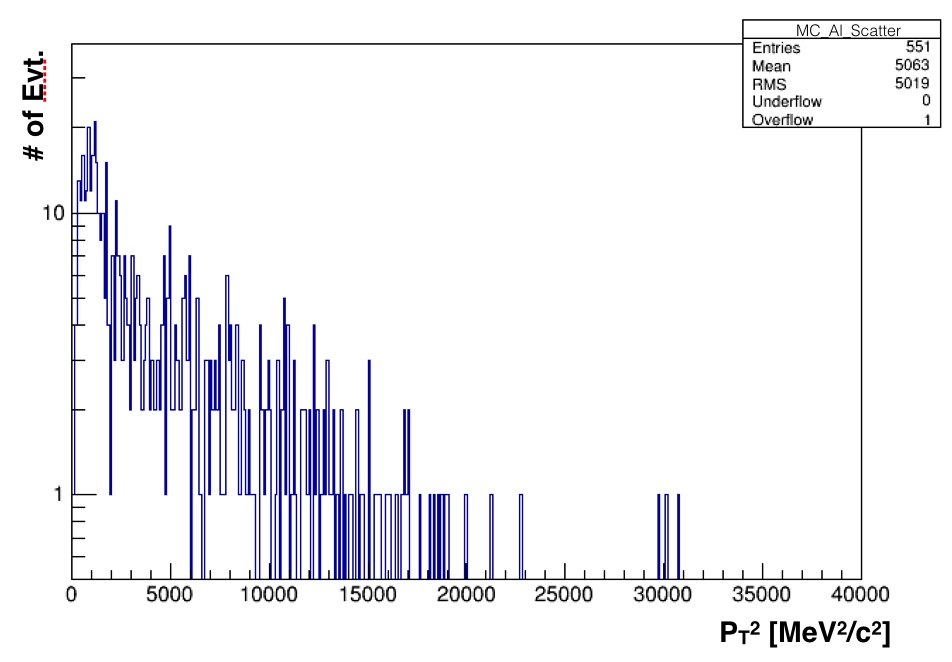,height=2.5in}
\caption{The prelimary study of scattered $K_L$ from the MC samples.}\label{fig:MCpt}
\end{center}
\end{figure}

The MC study is performed by inserting two tags sandwiching the Al target and track the momentum changes to determine the occurrence of $K_L$ scattering. A spectrum of $K_L$ beam profile is considered in the MC simulation and the preliminary study shows the a 1.17\% generation rate of scattered $K_L$ interacting with the Z0 Al target, which corresponds to 0.022\% for the vacuum window. The scattered $K_L$ is determined by the observation in changes of the $K_L$ transverse momentum between two tags, the distribution of the $K_L$ momentum change is shown in Fig. \ref{fig:MCpt}.  

\subsection{Al Target Data Study}
\begin{figure}[htb]
    \centering
    \begin{subfigure}[t]{0.5\textwidth}
        \centering
        \epsfig{file=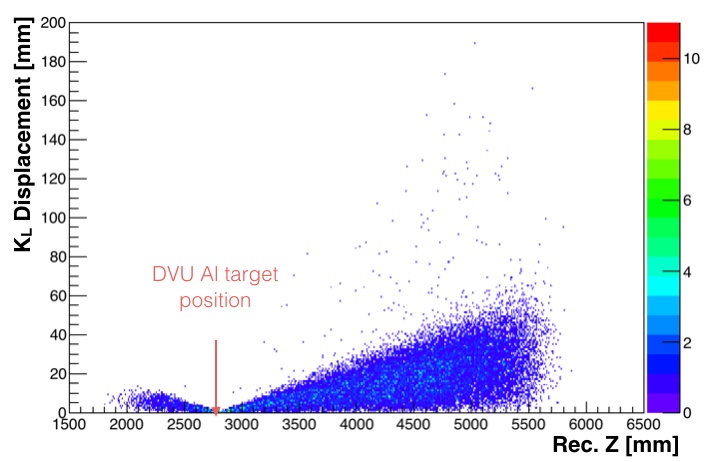,width=\textwidth}
    \end{subfigure}%
    ~ 
    \begin{subfigure}[t]{0.5\textwidth}
        \centering
         \epsfig{file=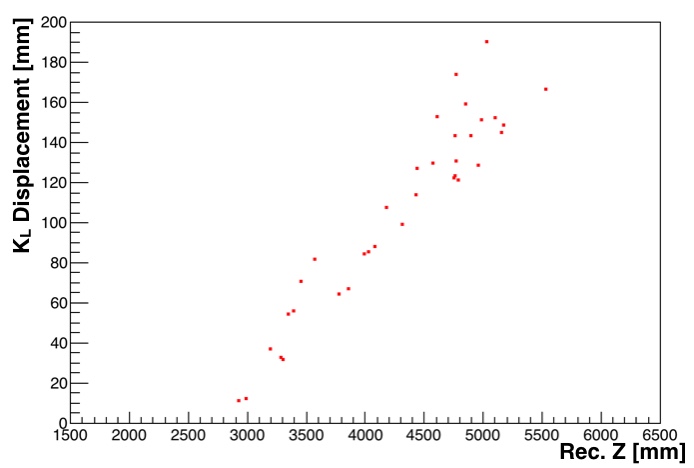,width=\textwidth}
    \end{subfigure}
    \caption{$K_L$ beam profile using the DVU Al target as the source of the $K_L$. The left plot shows the distrubtion of the $K_L$ off-axis displacement with respect to the decay z vertex position. The right plot is the left plot after applying the CoE radius cut at 200~mm. }\label{fig:DUVscatter}
\end{figure}

$K_L\rightarrow\pi^0\pi^0\pi^0$ data collected from the DVU Al target data in 2017 June run were used to study the $K_L$ beam profile. Instead of projecting the CoE back to the gold target, DVU Al target was used as the source of $K_L$ generation. Figure \ref{fig:DUVscatter} shows the distribution of the reconstructed $K_L$ displacement. We observe some potential scattered $K_L$ which has larger displacements than most of the $K_L$ in the left plot. After applying a CoE radius threshold of 200~mm, which is similar to the online CoE threshold inside the data acquisition system, some off-axis $K_L$ with large displacements from the beam axis remain. The signal region for the $K_L\rightarrow\pi^0\nu\bar{\nu}$ study in the KOTO experiment is between 3000~mm to 5000~mm. Therefore, the scattered $K_L$ in the right plot located within the signal region will consequently appear to have large transverse momentum and contribute to the background of $K_L\rightarrow\pi^0\nu\bar{\nu}$ signal.

\section{Summary}
The preliminary study on the MC samples confirmed the existence of the scattered $K_L$ and the DVU Al target data verifies the contribution of scattered $K_L$. The $K_L$ beam profile provided insights on background contributions to the signal. More statistics and studies on both Al targets are required to estimate the background contribution of the scattered $K_L$ and the work is in progress. 

\section{Acknowledgement}
This work is supported by DoE grant DE-SC0007859.

\end{document}